\documentstyle[12pt,aasms4]{article}
\begin{document}

\title{ The dynamic energy source of the Sun and 
the duplicity of the stellar energy production}

\author{Attila Grandpierre}

\affil{Konkoly Observatory, P.O. ~Box ~67, H--1525,
Budapest, Hungary \\
electronic mail: grandp@ogyalla.konkoly.hu}

\begin{abstract}
Some possible ways of the energy production with fusion reactions in the 
Sun was explored theoretically in the first half of this century. Nowadays it 
is a standard view that the Sun produces its energy on a uniform level. I 
point out, that in the stellar and solar energy production a dynamic energy 
source is necessarily present behind the uniform one, and generates a direct 
connection between the core and the surface layers through tunnels. 
\end{abstract}

\section{Introduction}
The standard solar model (SSM, [1,14]) states that the proton-proton 
cycle gives the 96.5\% of the total solar energy production, the CNO cycle gives 
only 1,5\% and the neutrinos make the remaining 2\%. Nevertheless, the first 
experimental check of the SSM measured only the third of the predicted solar 
neutrino flux. One could think that the plausible cause could be that the energy
production of the solar core is less than predicted. But it is hard to construct a 
solar model with a lower central temperature, which gives back the neutrino 
fluxes and the helioseismological tests as well, as the SSM can do. Moreover, 
the change of the standard solar model to a non-standard one would have 
consequences to the stellar evolution theories. Yet, the standard solar model 
seems to fit the best to the observational tests of the stellar evolution theories. So 
it seemed the best way to search the solution of the solar neutrino problem not in 
the astrophysics, but in the physics of the neutrinos, which would affect only the 
properties of the neutrinos on their way from the Sun to the Earth, without 
changing the physics in the Sun itself.

	The fundamental characteristic of stellar energy production is the 
temperature-sensitivity [2,3]. The larger the mass of a star, 
the higher the temperature in its core, and therefore, the faster the nuclear 
reactions. That is the cause why the giant stars have much shorter lifetime than 
the dwarfs. The temperature-sensitivity can generate instabilities in the stellar 
cores. At higher temperatures the nuclear reactions proceed much faster, which 
produces much more heat, generating still higher temperature and so on. The 
whole star would explode in the absence of a stabilising agent. That stabilising 
agent is the gravitation for the star as a whole. The virial theorem shows that the 
heat energy is less than the gravitational, therefore the coefficient of the heat 
expansion of the star is negative. This means that when the stellar core becomes 
hotter, it expands, and the volume expansion against gravity cools down the star 
more effectively than the nuclear reactions heat it. This is the generally accepted 
argument for the idea that the stellar cores are thermally stable. But this 
statement is not valid for local heating, since in a local volume the gravity do not 
have such a significant effect, and, at the same time, the local heating time is 
much shorter, and so the thermo-nuclear instability is present. The calculations 
show that the time-scale of the thermonuclear expansion is $10^{-5}s$ [2], much 
shorter than that of the volume expansion.

A principal theoretical difficulty arises at the origin of convection. We 
know that at most stars convection is present as a flow setting up by high 
temperature differences. But for the onset of convection it is also necessary that 
the initial perturbations has to be present. Since the convective cells has a 
characteristic size of some 1000 km, this would imply that initial perturbations 
should develop in a huge macroscopic volume. Nevertheless, atomic collisions 
can never generate macroscopic perturbations, since the collisions act to decay 
the fluctuations and since the convective zone before the convection sets up is in 
 a radiative equilibrium and radiation also acts to smooth any occasionally 
developing fluctuation [4,5]. Therefore, a mechanism has to exist 
which is able to develop macroperturbations from below the convective zone. I 
was led to investigate the possible instabilities developing in the solar core in 
order to find a mechanism to generate the macroperturbations. 
It also seemed that the apparent fluctuations of the Homestake neutrino 
flux have significantly larger amplitude than the observational errors plus 
statistic fluctuations would allow it. This circumstance again pointed to the 
existence of an instability in the energy producing solar core. 

My calculations on 
the thermonuclear time-scales were strengthened by the arguments of 
Zel'dovich et al. [6] who has shown that the dissipative mechanism are too 
ineffective to stabilise against the local thermonuclear instability. Nevertheless, 
they did not attempt to find the way out of the dilemma of the local instability 
and global stability of the stellar energy producing regions.
My calculations led to the picture that the fundamental local thermonuclear instability of stellar cores generates 
"hot bubbles", plumes which has very different nature from the convective 
flows. In the case of the Earth, the plumes from the edge of the core have 50-100 
larger speeds than the convective flows litospheric plates. The diameter of the 
'hot spots' of the Earth are around 200-300 meters and in this areas the heat 
upflow in the mantle is five times larger than elsewhere. This is why they are 
called 'hot spots', even when they do not seem to be related to any surface 
vulcano. These hot spots has a long lifetime of some ten or hundred millions of 
years, but they did not seem to participate in the plate-tectonic movements, 
which show their deep origin and rigid rotation. The solar 'hot spots' show as 
well anomalous activity, higher temperature, and rigid rotation. The parallel 
phenomena suggest that the flow in them also could be higher then elsewhere 
and that their material originates from significant depths.
Beside the macro-instability producing the hot bubbles from the core, the 
local thermonuclear instability produces also microscopic effects. The stochastic 
atomic collisions can produce thermal instabilities, but since their size do not 
reach the critical threshold given by the actual Rayleigh number, which depends 
strongly on the size of the perturbation [5], they will decay. 
Nevertheless, they will be continuously re-generated and because of the 
temperature-sensitivity the effect of the microinstability will be that thermal 
inequilibrium will develop. Kaniadakis, Lavagno and Quarati [18,19] calculated 
some effects of the modified Maxwell-distribution to the solar structure and the 
neutrino spectrum. Their result is similar to that of the microinstabilities: they act to generate different temperatures of the ions and electrons, to increase the central temperature, and that they effect to change the neutrino spectrum 
towards the observed one. In this way, the effect of the micro-and macro-
instabilities of the solar core seem to compensate each other in respect to the 
solar structure, and modify the neutrino production towards a better agreement 
with the observations. This compensating effect could be a reason why the 
standard solar model seems still close to the helioseimological results.

In between 1994.10.12 - 1995.10.04 the GALLEX group [10] made 14 
measurements. Their results, the GALLEX-III gives the lowest value yet 
measured, 41\% of the SSM. This value is significantly lower than the 56\% 
threshold which is compatible with the idea that the solar energy is supplied 
through proton-proton cycle. Since the error-bars sum up only to 10\%, this result 
is the first measurement to show that there has to be an additional energy source 
of the Sun besides the proton-proton (and CNO) cycle, which produces a 
significant part of the solar energy supply. This result cannot be regarded as a 
mere fluctuation, since in the same time interval the KAMIOKANDE-III 
measurements also has shown the lowest measured rate, 34\% of the SSM value [11]. 
The GALLEX rate $53.9 \pm 11$ SNU inevitably shows that it is not possible 
to produce all the solar luminosity with the proton-proton cycle, since in that 
case the minimum flux should be 87 SNU [7,8], which is 
above the $3 \sigma$ threshold. These new neutrino measurements suggest an $ 
8 \%$ decrease of temperature. But in this case an additional source of energy 
generation has to be present which has to produce the missing 28\% of the solar 
luminosity. I suggest that this new type of energy production is produced in 
local thermonuclear runaways [2]. So we have to abandon the 
luminosity constraint in the context of a steady and hydrogen-burning solar core, 
since an additional new type of energy source is apparent. 

Another fact is that the KAMIOKANDE is the only detector which is 
sensitive to neutral currents. Therefore, muon and tau neutrinos 
produced in the hot bubbles will be detected by the KAMIOKANDE and so one
has to subtract their contribution from the observed rates. This circumstance 
offers another way to solve the apparent contradiction between the observations 
of the different neutrino detectors, and the beryllium neutrino problem. 

These factors reveal the caveat in the 
argument of Castellani et al. [7] against all astrophysical solution, and 
makes it possible to construct dynamic solar core models. The attack
of the paper is using all the data of the neutrino detectors, 
to present calculations showing a new possibility for the solutions
of the solar neutrino problems, and to accept the new results GALLEX-III and KAMIKANDE-III by their face values, showing that it gives another indication for a non-standard Sun. I use different effective temperatures for the different neutrino
productions. Haubold and Mathai [13] observed that the solar neutrino 
problem may have an astrophysical solution when the deviations of the 
actual neutrino temperatures from the SSM values are different for
the different neutrino sources.

\section{ General equations of the individual neutrino fluxes }

The basic equations describing the neutrino production of the quiet solar 
core in terms of the individual neutrino fluxes $\Phi_p$, $\Phi_{Be}$, 
$\Phi_{CNO}$ and $\Phi_B$, and the observed rates with the gallium-detector
$S_G$ and the chlorine-detector $S_C$ are 
\begin{eqnarray}
S_G=S_G(B)+G_p\Phi_p+G_{Be}\Phi_{Be}+G_{CNO}\Phi_{CNO}
\end{eqnarray}
\begin{eqnarray}
S_C=\Phi_{Be}+C_{CNO}\Phi_{CNO}+C_B\Phi_B+C_{pep}\Phi_{pep},
\end{eqnarray}
where $C_i$ and $G_i$ are the detector sensitivities of the chlorine and 
gallium detectors, given by Table I. in [8].
Solving these equations for $\Phi_{Be}$ and $\Phi_p$, 
\begin{eqnarray}
\Phi_{Be}= (S_C-C_B\Phi_B-C_{CNO}\Phi_{CNO} -
C_{pep}\Phi_{pep})/C_{Be}
\end{eqnarray}
\begin{eqnarray}
\Phi_p=(S_G-G_{Be}S_C/C_{Be}+ 
\alpha_B\Phi_B+\alpha_{CNO}\Phi_{CNO})/G_p
\end{eqnarray}
\begin{eqnarray}
\alpha_i = G_{Be}/C_{Be}C_i - G_i.
\end{eqnarray}	
For the individual fluxes their time dependence may be approximated as
\begin{eqnarray}
\Phi_B=R(K)\Phi_B(SSM) = T_B^{24.5}\Phi_B(SSM),
\end{eqnarray}
where $T_B$ (and later on all the $T_i$ temperatures) are dimensionless 
temperatures, normalised to the standard value, 
$T_B=T_B(average, actual)/T_B(average, SSM)$.
\begin{eqnarray}
\Phi_{Be}=R_{Be}\Phi_{Be}(SSM)= T_{Be}^{11.5}\Phi_{Be}(SSM)
\end{eqnarray}
\begin{eqnarray}
\Phi_{CNO}=R_{CNO}\Phi_{CNO}(SSM)= T_{CNO}^{20} 
\Phi_{CNO}(SSM)
\end{eqnarray}
\begin{eqnarray}
\Phi_p=R_p\Phi_p(SSM)=T_p^4\Phi_p(SSM)
\end{eqnarray}
Now inserting these temperature-dependent equations into the basic equations
of the neutrino fluxes (3), (4), I obtain the two basic temperature-dependent
equations for the neutrino fluxes (neglecting the $pep$ fluxes):
\begin{eqnarray}
T_B^{24.5}C_B/C_{Be}\Phi_B(SSM)+T_{CNO}^{20}C_{CNO}/C_{Be}\Phi
_{CNO}(SSM)+T_{Be}^{11.5}\Phi_{Be}=S_C/C_{Be}
\end{eqnarray}
\begin{eqnarray}
T_B^{24.5}\alpha_B\Phi_B(SSM)+T_{CNO}^{20}\alpha_{CNO}\Phi_{CNO}
(SSM)-T_p^4G_P\Phi(SSM)=S_G-G_{Be}/C_{Be}S_C
\end{eqnarray}
Now I regarded $T_B$, $T_{Be}$ and $T_{CNO}$ as being equal with $T_c$, 
since they are all characteristic to temperatures of the different maximum 
neutrino productions, at $r=0.04, 0.06$ and 
$0.05R_{Sun}$, i.e. relatively close sites. Nevertheless, I allowed 
$T_p$ to be different, because $T_p$ is characteristic for a region around 
$r=0.10R_{Sun}$,
which may be regarded as being a site not too close to the above three.
In this way I have only 
two unknowns to be determined, $T_c$ and $T_p$, and I have two equations
for them. 

Using the observed time-averaged values of the neutrino fluxes, 
$S_G=69.7SNU$ [10], $S_C=2.56 SNU$ [13],
the standard neutrino fluxes from [14] are 
$\Phi_B(SSM)=5.71 \times 10^6cm^{-2}s^{-1}$,
$\Phi_{CNO}(SSM)= 1.1 \times 10^9 cm^{-2}s^{-1}$, 
$\Phi_{Be}(SSM)= 0.47 \times 10^{10}cm^{-2}s^{-1}$,
$\Phi_p(SSM)=5.71 \times 10^{10} cm^{-2}s^{-1}$, 
$C_{Be}=0.24 \times 10^{-9}$,
$C_B=1.09 \times 10^{-6}$, $G_{Be}=7.32 \times 10^{-9}$, 
$G_B=2,43 \times 10^{-6}$, $C_{CNO}=0.40 \times 10^{-9}$, 
$G_{CNO}=8.67 \times 10^{-9}$, $G_p= 1.67 \times 10^{-9}$,
the solutions are $T_c=0.95$ and $T_p=0.96$. 
The obtained results show that no beryllium-neutrino problem arise, and the 
temperatures are remarkably close to the most recent seismological solar models 
[9]. The determination of the temperatures in the solar core with the above 
equations present a very sensitive method for the temperatures of the different 
layers of the solar core, as their one percent variation already
leads to values incompatible with equations (10) and (11). In this way the 
presented general calculation of the solar core temperatures remarkably do not
lead to any solar neutrino problem, as it contains the basic physics
and so it has definite consequences for the other neutrino fluxes which
are not included directly. For example, in the case of a constant
solar core a boron neutrino flux from the quiet solar core $\Phi_B=0.28$ 
(or $0.37$) $\Phi_B(SSM)$ is required
with $T_B=0.95$ (or $T_B=0.96$). This means that the remaining part of the 
neutrinos, as observed by the KAMIOKANDE, do not originate from the quiet 
core, but from the hot bubbles. The KAMIOKANDE is the only detector, which 
is 
sensitive to neutral currents. Therefore, heavy neutrinos produced outside of 
the proton-proton cycle may be detected by the KAMIOKANDE and so we 
should subtract their contribution from the observed rates. There is no such
a problem as the problem of beryllium neutrinos; instead, the basic
equations of the neutrino fluxes state clearly that the Kamiokande observes
neutrinos besides the boron neutrinos of the quiet solar core.

The results obtained for a static solar core present solid evidence for
a solar core being cooler than standard and, at the same time, it is also
indicated that there are selective deviations from the standard solar model 
in the different depths of the solar core. These results have a high relevance
in the study of the solar neutrino problem and in constructing realistic solar 
models. Having found such a sensitive tool for the study of the solar core
as the equations presented above, I applied these equations
to a solar core varying in time as well. 

In case of solar activity minimum, I can use $S_C(min)=4.1SNU$ [17]
and $S_G(min)=53.9SNU$ [10]. With these values (10) gives
$T_c=0.973$ and substituting this value to (11) an unphysical value of 
$T_p=2.17$
arises. The cause of this discrepancy could be i.) observational errors in
$S_C$ and $S_G$, or ii.) an unidentified flux contributes to the Homestake
detection rates, for which the GALLEX is less sensitive. Regarding ii.), 
it is known that
in the Homestake the contribution of the intermediate energy neutrino fluxes 
are around $30 \%$ while they constitute only around $2 \%$ in the gallium
detectors. Therefore, the time-dependence of the solar energy and neutrino
production indicates that in the solar activity minimum the yet unidentified 
flux is supplied by intermediate energy neutrinos, produced by the hot 
bubbles. 

In case of solar activity maximum, $S_C(min)=2.3 SNU$ [13]. With 
$S_G(max)=
79SNU$ [10] the derived values are $T_c=0.945$ and so $T_p=0.77$. 
Regarding 
such a large deviation as unphysical points to the presence of a yet unnoticed 
neutrino flux present around solar activity maximums. This additional neutrino 
flux $\Phi_b(max)$ has to give a term besides the boron neutrino fluxes in
(10) and (11) if we want a physically consistent description of the solar core
using the observed neutrino fluxes. In this way I identified another yet 
unrecognised
physical process present in the solar core, being active around solar
activity maximums, producing high-energy neutrinos. 

It is interesting, that a puzzling difference is observed between the 
frequency shifts of the even $l$ = 0, 2 and odd $l$ = 1, 3 modes [15]. These 
significant differences indicate a "sandwich" structure of the Sun, a 
coupling between the different depths at the very neighbourhood of the 
centre. A similar phenomenon may occur if there is a direct transfer from 
a central region to an outer layer.

Another interesting point, that the flares at the solar surface do show 
central 'tunnels' between their footpoints to the loop-tops [16], indicating 
that they set up as the consequence of subphotospheric mass outflows. This 
"unexpected" phenomenon was predicted ten years ago on a purely theretical
basis of the convective flare theory [17] declaring a direct connection of the
solar core with the surface through isolated "channels" or "volcanic funnels".

%
%
\section{Conclusions}

The main result of the presented calculations is that all the neutrino 
measurements are indicative and quite compatible with the theoretical 
result that a new type of energy production mechanism is active in the 
solar core. 
Our calculations outline its physical nature and suggest that at solar 
minimum it produces intermediate energy CNO neutrinos and at solar maximum 
it 
possibly contributes to the flare effect in the high energy neutrino fluxes. 
These predictions can be proven with future measurements of the solar 
neutrino spectra.

The obtained results show that for the time-averaged values of the solar core
the Sun shows a temperature $0.96 T(SSM)$ around $r=0.10R_{Sun}$ while 
at the deeper layers around $r=0.05R_{Sun}$ the temperature has a different 
value of $0.95T(SSM)$. This result can not be regarded
as marginal since the equations describing the production of the individual
neutrino fluxes are highly sensitive to the temperatures and $T_c=T_p=0.96$ 
would lead to larger than three $\sigma$ deviations from the observed values, 
to $S_C=3.19 SNU$ and $S_G=104.8 SNU$. 

A scheme of the solar structure is derived, which has a definite 
suggestion that below 0.10 solar radius the standard solar model is to be 
replaced by a significantly cooler and varying core. Nevertheless, it is 
indicated that the compensating effect of the thermonuclear 
micro-instabilities effectuating an increase in the central temperatures 
and a parallel decrease in the neutrino fluxes [18, 19] balances the cooling 
effect. 

Another prediction of the dynamic solar core model is that the chemical 
composition of the solar wind varies significantly with the phase of the 
activity cycle, more enhanced in heavy elements near to maximum. SOHO can 
test this prophecy. Crooker published results showing this effect: 
"the proton temperature and velocity closely anti-correlates with the 
electron density and temperature in the solar wind" [20].

The discovery of the dynamic energy source of the stars has a significance 
in relation to the world-view of science as well. The dynamic energy production 
is very sensitive to the effects of the environment, to the week tidal effects, and 
so it is able to couple such far branches of science as the celestial mechanics, the 
nuclear astrophysics and the stellar activity phenomena. The conclusion that the 
planets participate in the regulation of the solar energy production show that the 
Sun cannot be regarded as a closed system but an open one in its most 
fundamental nature.

\section{Acknowledgement}

The work is supported by the Hungarian Scientific Research Foundation OTKA
under No. T 014224.

\end{document}